%% file: rhumo.tex
\documentclass[10pt,journal,compsoc]{IEEEtran}
\ifCLASSOPTIONcompsoc
  \usepackage[nocompress]{cite}
\else
  \usepackage{cite}
\fi

\ifCLASSOPTIONcompsoc
  \usepackage[nocompress]{cite}
\else
  \usepackage{cite}
\fi
\ifCLASSINFOpdf
\else
\fi

\usepackage{graphicx}
\usepackage{times,amsmath,epsfig}
\usepackage{float}
\usepackage{amssymb}
\usepackage{multirow}
\usepackage{array}
\usepackage{chngpage}
\usepackage{subfigure}
\usepackage{color}
\usepackage[ruled,linesnumbered,boxed]{algorithm2e}
\usepackage{amsthm}
\usepackage{enumitem}

\setenumerate[1]{itemsep=0pt,partopsep=0pt,parsep=\parskip,topsep=5pt}
\setitemize[1]{itemsep=0pt,partopsep=0pt,parsep=\parskip,topsep=5pt}
\setdescription{itemsep=0pt,partopsep=0pt,parsep=\parskip,topsep=5pt}

\theoremstyle{plain}
\newtheorem{definition}{Definition}
\newtheorem{lemma}{Lemma}
\newtheorem{theorem}{Theorem}

\begin{document}
%
\title{\emph{r}-HUMO: A Risk-aware Human-Machine Cooperation Framework for Entity Resolution with Quality Guarantees}
%
%
%
%
\author{Boyi~Hou,
        Qun~Chen,
        Zhaoqiang~Chen,
        Youcef~Nafa
        and~Zhanhuai~Li
\IEEEcompsocitemizethanks{\IEEEcompsocthanksitem 1. School of Computer Science, Northwestern Polytechnical University.\protect\\
2. Key Laboratory of Big Data Storage and Management, Northwestern Polytechnical University, Ministry of Industry and Information Technology\protect\\
127 West Youyi Road, Xi'an Shaanxi, China\protect\\
E-mail: \{ntoskrnl@mail., chenbenben@, chenzhaoqiang@mail., youcef.nafa@mail., lizhh@\}nwpu.edu.cn}
\thanks{Manuscript received; revised.}}

%
%

\markboth{}%
{Boyi Hou \MakeLowercase{\textit{et al.}}: {\em r}-HUMO: A Risk-aware Human-Machine Cooperation Framework for Entity Resolution with Quality Guarantees}
%



\IEEEtitleabstractindextext{
\begin{abstract}
  Even though many approaches have been proposed for entity resolution (ER), it remains very challenging to enforce quality guarantees. To this end, we propose a \emph{r}isk-aware HUman-Machine cOoperation framework for ER, denoted by {\em r}-HUMO. Built on the existing HUMO framework, {\em r}-HUMO similarly enforces both precision and recall guarantees by partitioning an ER workload between the human and the machine. However, \emph{r}-HUMO is the first solution that optimizes the process of human workload selection from a risk perspective. It iteratively selects human workload by real-time risk analysis based on the human-labeled results as well as the pre-specified machine metric. In this paper, we first introduce the {\em r}-HUMO framework and then present the risk model to prioritize the instances for manual inspection. Finally, we empirically evaluate {\em r}-HUMO's performance on real data. Our extensive experiments show that {\em r}-HUMO is effective in enforcing quality guarantees, and compared with the state-of-the-art alternatives, it can achieve desired quality control with reduced human cost.
\end{abstract}

\begin{IEEEkeywords}
Entity Resolution, Human-Machine Cooperation, Risk Analysis, Quality Guarantee.
\end{IEEEkeywords}}

\maketitle

\IEEEdisplaynontitleabstractindextext

%
\IEEEpeerreviewmaketitle

\input{introduction}
\input{relatedwork}
\input{preliminaries}
\input{framework}
\input{riskanalysis}
\input{experiment}

\vspace{-0.05in}
\section{Conclusion} \label{sec:Conclusion}

In this paper, we have proposed a risk-aware human-machine cooperation framework, \emph{r}-HUMO, for entity resolution with quality guarantees.
Different from the existing HUMO framework, {\em r}-HUMO takes advantage of the manually-labeled results to measure the risk of pairs being mislabeled by machine, thus can effectively reduce required manual work. Our extensive experiments on real data have also validated the efficacy of {\em r}-HUMO.

  For large workload, crowdsourcing may be the only feasible solution for human verification. It is interesting to integrate \emph{r}-HUMO into the existing crowdsourcing platforms in future work. On the crowdsourcing platforms, monetary cost may be a more appropriate metric of human cost than the number of manually inspected pairs used in this paper. On the other hand, re-training the machine learning algorithm after each iteration of manual labeling can usually improve the overall performance of human and machine cooperation. It is a challenging task and deserves an independent investigation in future work.

\vspace{-0.1in}
\section*{Acknowledgments}
This work is supported by the Ministry of Science and Technology of China, National Key Research and Development Program (2016YFB1000703), National Natural Science Foundation of China (61332006, 61732014, 61672432, 61472321 and 61502390).

\ifCLASSOPTIONcaptionsoff
  \newpage
\fi



%

\vspace{-3\baselineskip}

\end{document}

%% file: introduction.tex
\IEEEraisesectionheading{\section{Introduction}\label{sec:introduction}}

   \IEEEPARstart{E}{ntity} resolution (ER) usually refers to identifying the relational records that correspond to the same real-world entity. A challenging task due to incomplete and dirty data, ER has been extensively studied in the literature~\cite{christen2012data, elmagarmid2007duplicate, fellegi1969theory, fan2009reasoning, li2015rule, singh2017generating, sarawagi2002interactive, christen2008automatic, mozafari2014scaling, li2017human, zhuang2017Hhike}. Unfortunately, it remains very challenging to enforce quality guarantees on ER. The approach based on active learning~\cite{arasu2010active, bellare2012active} can maximize recall while ensuring a pre-specified precision level. More recently, a HUman-Machine cOoperation framework \cite{humodemo, humo}, denoted by HUMO, has been proposed to enforce more comprehensive quality guarantees at both precision and recall fronts. HUMO enables a flexible mechanism for quality control by partitioning an ER workload between the human and the machine. It automatically labels easy instances by the machine while assigning more challenging ones to the human. For instance, given a metric of record pair similarity, the pairs with high or low similarities can be automatically labeled by the machine with high accuracy. However, the pairs with medium similarities may require human inspection because labeling them either way by machine would introduce considerable errors. The optimization objective of HUMO is to minimize the required human cost given the user-specified precision and recall levels.

	HUMO measures the hardness of an ER instance pair by a pre-specified machine metric and performs human workload selection in batch mode. It first groups the pairs into subsets by their metric values and then assigns the subsets between the human and the machine. As a result, all the pairs with similar metric values in a subset would be either automatically labeled by the machine or manually labeled by the human. However, it can be observed that due to the limitation of machine metrics, even though two pairs have similar metric values, their risks of being mislabeled by the machine may be vastly different.

	In this paper, we investigate the problem of workload partition between the human and the machine from a risk perspective. Since human workload selection can be performed in an interactive manner, human input, which consists of the human-labeled pairs in our example, can be naturally used for risk analysis to prioritize pairs for human inspection. Our idea is to iteratively pick up more risky pairs from a given subset of pairs for human inspection such that the remaining pairs in the subset can achieve overall higher machine-labeling accuracy. With human effort spent on more risky pairs, the required human cost for quality guarantees can be effectively reduced. As HUMO, the proposed risk-aware framework, {\em r}-HUMO, is to some extent motivated by the success of the existing crowdsourcing solutions for ER~\cite{chai2016cost, li2016crowdsourced, wang2012crowder, vesdapunt2014crowdsourcing}. The work on crowdsourcing ER focused on how to make the human work effectively and efficiently on a given workload. HUMO and {\em r}-HUMO instead investigate how to partition a workload between the human and the machine such that a user-specified quality requirement can be met.
	
	The major contributions of this paper can be summarized as follows:
	
\begin{enumerate}
\item We propose a risk-aware human-machine cooperation framework for ER, \emph{r}-HUMO, which can enforce both precision and recall guarantees. It is the first solution that optimizes the process of human workload selection from a risk perspective.
\item We propose the technique of risk analysis for iterative human workload selection. We present the risk model based on modern portfolio investment theory to prioritize ER pairs for human inspection;
\item We conduct an empirical study on the performance of \emph{r}-HUMO by extensive experiments on real data. Our experimental results show that \emph{r}-HUMO is effective in enforcing quality guarantees, and compared with the state-of-the-art alternatives, it can achieve desired quality control with reduced human cost.
\end{enumerate}

   The rest of this paper is organized as follows: Section~\ref{sec:Related} reviews more related work. Section~\ref{sec:Preliminaries} introduces the problem and briefly describes the existing HUMO framework.  Section~\ref{sec:Framework} presents the \emph{r}-HUMO framework. Section~\ref{sec:riskmodel} describes the technique of risk analysis. Section~\ref{sec:Experiments} presents our empirical evaluation results.  Finally, Section~\ref{sec:Conclusion} concludes this paper with some thoughts on future work. 

%% file: relatedwork.tex
\section{Related Work} \label{sec:Related}

  As a classical problem in the area of data quality, entity resolution has been extensively studied in the literature~\cite{christen2012data, elmagarmid2007duplicate}. It can be performed based on rules~\cite{fan2009reasoning, li2015rule, singh2017generating}, probabilistic theory~\cite{fellegi1969theory, singla2006entity} or machine learning~\cite{sarawagi2002interactive, christen2008automatic, arasu2010active, bellare2012active}. Unfortunately, it remains very challenging to enforce quality guarantees on ER.

  The approach based on active learning~\cite{arasu2010active, bellare2012active} has been proposed to enforce the precision guarantee on ER. The authors of \cite{arasu2010active} proposed a technique that can optimize recall while ensuring a pre-specified precision level. The authors of~\cite{bellare2012active} proposed an improved algorithm to approximately maximize recall under the precision constraint. Compared with the work of~\cite{arasu2010active}, its major advantage is better label complexity. However, these techniques share the same classification paradigm with the traditional machine learning algorithms; hence they can not enforce comprehensive quality guarantees specified at both precision and recall fronts.

  The progressive paradigm for ER~\cite{whang2013pay, altowim2014progressive} has also been proposed for the application scenario in which ER should be processed efficiently but does not necessarily require to generate high-quality results. Taking a pay-as-you-go approach, it studied how to maximize quality given a pre-specified resolution budget. In~\cite{whang2013pay}, the authors proposed several concrete ways of constructing resolution ``hints'' that can then be used by a variety of existing ER algorithms as a guidance for which entities to resolve first. In~\cite{altowim2014progressive}, the authors studied the more complicated problem of relational ER, in which a resolution of some entities might influence the resolution of other entities. A similar iterative algorithm, SiGMa, was proposed in \cite{lacoste2013sigma}. It can leverage both the structure information and the string similarity measures to resolve entity alignment across different knowledge bases. There also exist some interactive systems~\cite{Chu2015Katara, Elmagarmid2014NADEEF} that take advantage of knowledge bases or specific user input to achieve improved efficiency and quality for data cleaning. Unfortunately, these techniques have been built on machine computation; hence they can not be applied to enforce quality guarantees either.
	
  It has been well recognized that pure machine algorithms may not be able to produce satisfactory results in many practical scenarios~\cite{li2016crowdsourced}. Therefore, many researchers ~\cite{mozafari2014scaling, chai2016cost, wang2012crowder, vesdapunt2014crowdsourcing, Gruenheid2012Crowdsourcing, Getoor2012Entity, Chu2015Katara, Firmani2016Online, whang2013question, gokhale2014corleone, wang2015crowd, verroios2017waldo} have studied how to crowdsource an ER workload. In~\cite{wang2012crowder}, the authors studied how to generate Human Intelligence Tasks (HIT), and how to incrementally select the instance pairs for human inspection such that the required human cost can be minimized. In~\cite{whang2013question}, the authors focused on how to select the most beneficial questions for the human in terms of expected accuracy. More recently, the authors of~\cite{chai2016cost} proposed a cost-effective framework that employs the partial order relationship on instance pairs to reduce the number of asked pairs. Similarly, the authors in~\cite{verroios2017waldo} provided a solution to take advantage of both pairwise and multi-item interfaces in a crowdsourcing setting. The authors of~\cite{Anja2018Cost} studied how to balance cost and quality in crowdsourcing. Considering the diverse accuracies of workers across tasks, the authors of~\cite{Fan2015iCrowd} proposed an adaptive crowdsourcing framework that assigns the tasks based on worker accuracy estimation. In~\cite{Firmani2016Online}, the authors proposed an online crowdsourcing platform based on oracle. While these researchers addressed the challenges specific to crowdsourcing, we instead investigate a different problem in this paper: how to partition a workload between the human and the machine such that a user-specified quality requirement can be met. Since the workload assigned to the human by {\em r}-HUMO can be naturally processed by crowdsourcing, our work can be considered orthogonal to the existing work on crowdsourcing. It is interesting to investigate how to seamlessly integrate a crowdsourcing platform into {\em r}-HUMO in future work.

  The {\em r}-HUMO framework is built on the recently proposed HUMO framework~\cite{humodemo, humo}, which can enforce quality guarantees at both precision and recall fronts. The general idea of HUMO and {\em r}-HUMO was similar to the Fellegi-Sunter theory of record linking \cite{fellegi1969theory}, which also proposed to divide an ER workload into three parts based on match probability. HUMO however proposed the effective algorithms to divide an ER workload and estimate the match probability of machine workload for the quality guarantees specified at both precision and recall fronts. The {\em r}-HUMO framework represents a major step forward in that it is the first solution to optimize the process of human workload selection from a risk perspective. Instead of selecting human workload in batch mode purely based on a pre-specified machine metric as HUMO does, {\em r}-HUMO performs real-time risk analysis on the manually labeled results for the purpose of reducing the required human cost.

%% file: preliminaries.tex
\section{Preliminaries} \label{sec:Preliminaries}

\subsection{Problem Definition}

   Entity resolution reasons about whether two records are equivalent. Two records are deemed to be equivalent if and only if they correspond to the same real-world entity. We call a pair an {\em equivalent} pair if and only its two records are equivalent; otherwise, it is called an {\em inequivalent} pair. An ER solution labels each pair in a workload as {\em matching} or {\em unmatching}. As usual, we measure the quality of an ER solution by the metrics of precision and recall. Precision denotes the fraction of equivalent pairs among the pairs labeled as {\em matching}, while recall denotes the fraction of correctly labeled equivalent pairs among all the equivalent pairs.

\begin{table}[!h]
\centering
\caption{Frequently Used Notations.}
\label{tb:notations}
\begin{tabular}{|l|l|}
\hline
 Notation & Description \\ \hline
 $D$        &  an ER workload consisting of instance pairs \\ \hline
 $D_i$, $D_+$, $D_-$, $D_H$ & subsets of $D$ \\ \hline
 $S$, $S_i$ & a labeling solution for $D$ \\ \hline
 $d$, $d_i$ & an instance pair in $D$ \\ \hline
 $TN(D_i)$   & the total number of pairs in $D_i$ \\ \hline
 $EN(D_i)$  & the number of equivalent pairs in $D_i$ \\ \hline
 $EP(D_i)$ & the proportion of equivalent pairs in $D_i$ \\ \hline
 $f$, $f_i$ & a feature of instance pair \\ \hline
 $F$, $F_i$ & a feature set \\ \hline
 $D_f$      & the set of instance pairs with the feature $f$ \\ \hline
\end{tabular}
\end{table}

  For presentation simplicity, we summarize the frequently used notations in Table~\ref{tb:notations}. Formally, we define the problem of entity resolution with quality guarantees~\cite{humodemo, humo} as follows:

\begin{definition}
\label{problemsetting}
{\bf [Entity Resolution with Quality Guarantees].}  Given a set of instance pairs, $D=\{d_1, d_2, \cdots, d_n\}$, the problem of entity resolution with quality guarantees is to give a labeling solution $S$ for $D$ such that with the confidence level of $\theta$, $precision(D,S)\geq\alpha$ and $recall(D,S)\geq\beta$, in which $\alpha$ and $\beta$ denote the user-specified precision and recall levels respectively.
\end{definition}

\subsection{The HUMO Framework} \label{subsec:HUMO}

\begin{figure}[h]
\centering
\includegraphics[width=1.0\linewidth]{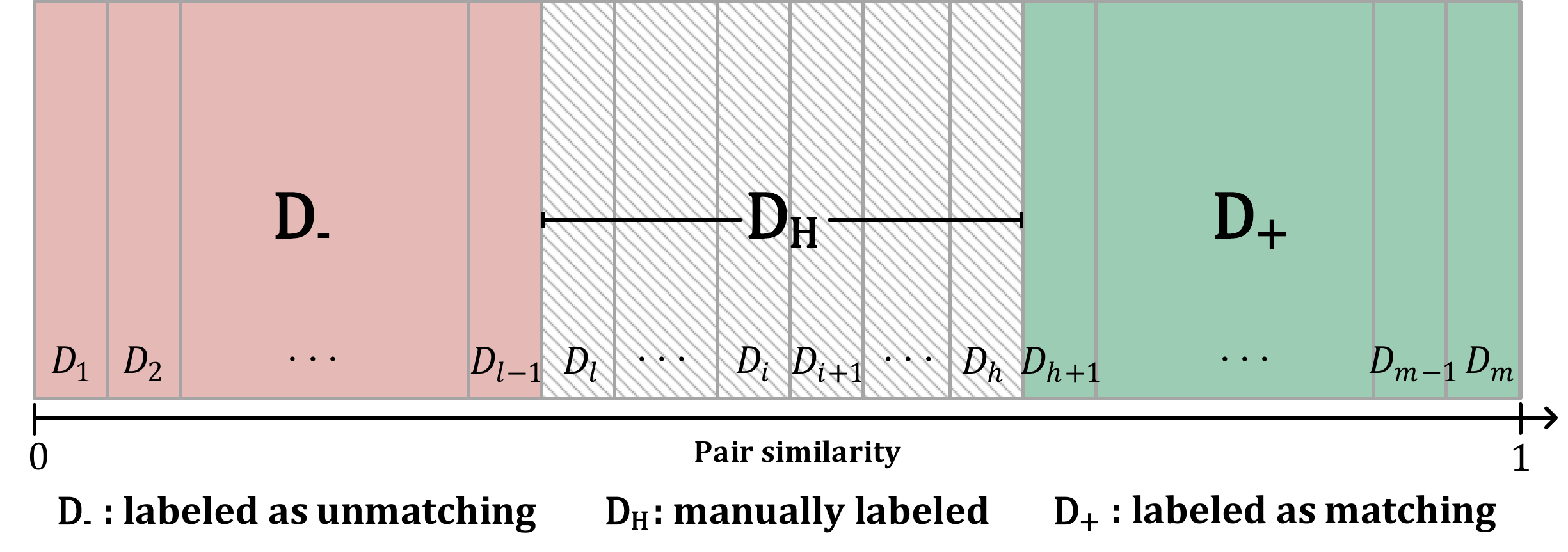}
\caption{HUMO Framework.}
\label{fig:humo}
\end{figure}

  The HUMO framework is shown in Fig.~\ref{fig:humo}. Given a workload, $D$, HUMO first groups its pairs into unit subsets (denoted by $D_i$ in the figure) by a machine metric (e.g., pair similarity or match probability), and then partitions the unit subsets into three disjoint sets, $D_-$, $D_H$ and $D_+$.  HUMO assumes that the given machine metric satisfies the monotonicity of precision, which statistically states that the higher (or lower) metric values a set of pairs have, the more probably they are equivalent pairs. HUMO enforces the precision and recall guarantees by automatically labeling $D_-$ and $D_+$ as {\em unmatching} and {\em matching} respectively, and assigning $D_H$ to the human for manual inspection.
	
	 The monotonicity assumption of precision underlies the effectiveness of HUMO's workload partitioning strategy between the human and the machine. However, HUMO never needs to expect that the monotonicity assumption can be \emph{strictly} satisfied on real data. Instead, it only assumes that provided with a reasonable machine metric, monotonicity of precision is usually a statistical trend on real data. It is also worthy to point out that in a similar way, the monotonicity assumption of precision underlies the effectiveness of the existing machine classification metrics for ER. HUMO is effective provided that the given machine metric satisfies the monotonicity assumption of precision. However, for presentation simplicity, we use pair similarity as the example of machine metric in this paper.

Given an ER workload, $D$, the quality of a HUMO solution, $S$, can be estimated by reasoning about the lower and upper bounds of the number of equivalent pairs in $D_-$, $D_H$ and $D_+$. In Figure.~\ref{fig:humo}, the lower bound of the achieved precision level can be represented by
\begin{equation}
   precision_L(D,S)=\frac{EN_L(D_+)+EN_L(D_H)}{TN(D_+)+TN(D_H)},
\label{eq:precision-bound}
\end{equation}
in which $TN(\cdot)$ denotes the total number of pairs in a set and $EN_L(\cdot)$ denotes the lower bound of the total number of equivalent pairs in a set. Similarly, the lower bound of the achieved recall level can be represented by
\begin{equation}
  recall_L(D,S)=\frac{EN_L(D_+)+EN_L(D_H)}{EN_L(D_+)+EN_L(D_H)+EN_U(D_-)},
\label{eq:recall-bound}
\end{equation}
in which $EN_U(\cdot)$ denotes the upper bound of the total number of equivalent pairs in a set. \emph{In this paper, for the sake of presentation simplicity, we assume that the pairs in $D_H$ can be manually labeled with 100\% accuracy. However, it is worthy to point out that the effectiveness of HUMO does not depend on the 100\%-accuracy assumption. It can actually work properly provided that quality guarantees can be enforced on $D_H$. In the case that human errors are introduced in $D_H$, the lower bounds of the achieved precision and recall can be estimated based on Eq.~\ref{eq:precision-bound} and Eq.~\ref{eq:recall-bound} respectively.} Nonetheless, under the assumption that the human performs better than the machine in resolution quality, the best quality guarantees HUMO can achieve are no better than the performance of the human on $D_H$.
	
	As human work is usually more expensive than machine computation, HUMO aims to minimize the workload in $D_H$ while guaranteeing resolution quality. By quantifying human cost by the number of instance pairs in $D_H$, we define the optimization problem of HUMO as follows~\cite{humodemo, humo}:

\begin{definition}
\label{optimization}
{\bf [Minimizing Human Cost in HUMO].} Given an ER workload, $D$, a confidence level $\theta$, a precision level $\alpha$ and a recall level $\beta$, the optimization problem of HUMO is represented by
\begin{equation}
\begin{split}
& \mathop{\arg\min}_{S}{|D_H(S)|} \\
& subject \quad to \quad P(precision(D,S)\geq\alpha)\geq\theta , \\
& \hspace{0.7in} P(recall(D,S)\geq\beta)\geq\theta ,
\end{split}
\label{eq:minimization}
\end{equation}
in which $S$ denotes a labeling solution, $D_H(S)$ denotes the set of instance pairs assigned to the human by $S$, $precision(D,S)$ denotes the achieved precision level of $S$, and $recall(D,S)$ denotes the achieved recall level of $S$.
\end{definition}

\begin{figure}[h]
\centering
\includegraphics[width=1.0\linewidth]{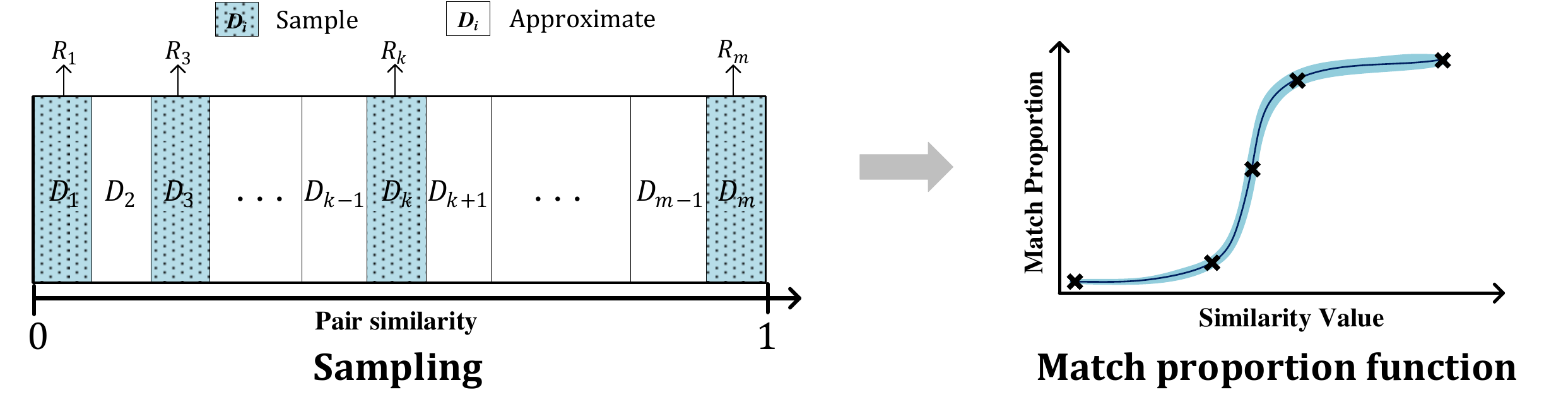}
\caption{Process of GPR.}
\label{fig:gpr}
\end{figure}

    The optimization problem as defined in Eq.~\ref{eq:minimization} is challenging because the proportions of equivalent pairs in $D_+$ and $D_-$ are unknown, thus need to be estimated. There exist two types of approaches to minimize the size of $D_H$: one purely based on the monotonicity assumption of precision and the other one based on sampling~\cite{humodemo, humo}. They estimate equivalence proportion based on different assumptions. Between them, the sampling-based approach has been empirically shown to have superior performance. It first estimates the equivalence proportions of the unit subsets by sampling, and then identifies the minimal workload of $D_H$ by reasoning about the numbers of equivalent pairs in $D_-$ and $D_+$. The equivalence proportion of a unit subset, $D_i$, can be directly estimated by sampling or approximated by Gaussian Process Regression (GPR)~\cite{rasmussen2006gaussian}. The process of GPR is shown in Fig.~\ref{fig:gpr}. Assuming that the equivalence proportions of all the unit subsets have a joint Gaussian distribution, GPR can approximate their equivalence proportions by sampling only a fraction of them. Based on GPR approximation, HUMO estimates the lower and upper bounds of the numbers of equivalent pairs in $D_-$ and $D_+$ by aggregating their corresponding Gaussian distributions. It can therefore iteratively optimize the lower and upper bounds of $D_H$. More technical details of HUMO can be found at \cite{humodemo, humo}.

%% file: framework.tex
\section{The \emph{r}-HUMO Framework} \label{sec:Framework}
	
\begin{figure}[h]
\centering
\includegraphics[width=1.0\linewidth]{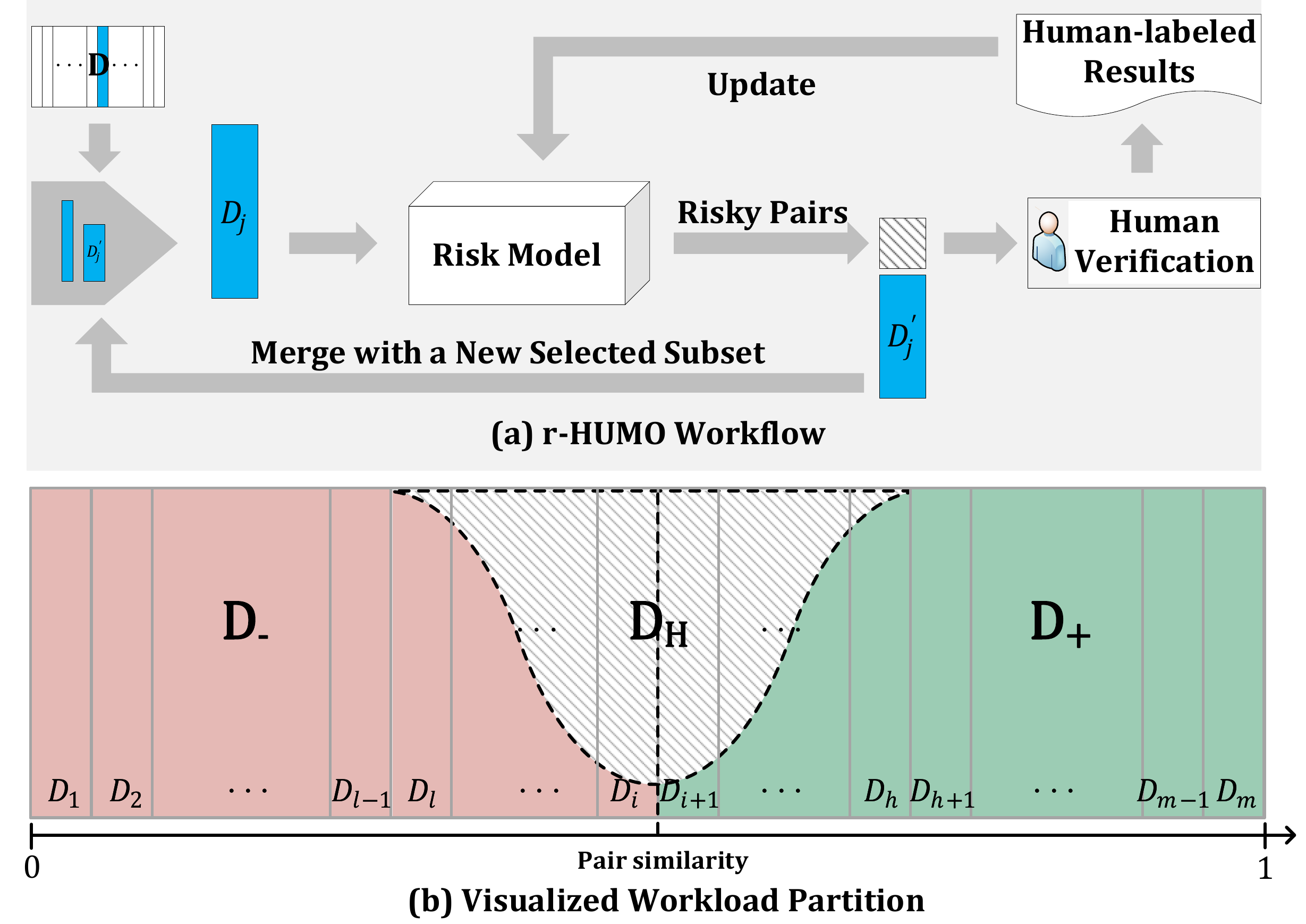}
\caption{{\em r}-HUMO Framework.}
\label{fig:r-HUMO}
\end{figure}

	 The \emph{r}-HUMO framework consists of two processes, human workload selection and risk analysis. The process of human workload selection picks out the pairs for manual inspection from a set of candidate pairs; the process of risk analysis estimates pair risk based on the human-labeled results. The procedure is invoked iteratively until the user-specified quality requirement is met. After each iteration, the set of candidate pairs is updated and pair risk is also re-estimated based on the updated set of human-labeled results. The \emph{r}-HUMO framework and its workflow are shown in Fig.~\ref{fig:r-HUMO}.
	
	 In the rest of this section, we first describe the basic real-time manner of human workload selection provided with a risk model, and then an alternative batch manner, which can significantly reduce the frequency of human-machine interaction. Finally, we present the technical details of quality assurance. However, the technique of risk analysis will be presented in the following section.
	
\vspace{-0.2in}

\subsection{Real-time Human Workload Selection}

  Suppose that $D$ has been divided into $m$ unit subsets with increasing metric values of pair similarity, $D=\{D_1,D_2,...,D_m\}$. Initially, we set $D_H=\emptyset$, $D_-=\{D_1,\ldots,D_i\}$, and $D_+=\{D_{i + 1},D_{i + 2},\ldots,D_m\}$. Since the pairs in $D_-$ would be automatically labeled as {\em unmatching}, the equivalence proportions of the subsets in $D_-$ are expected to be less than 0.5; similarly, the equivalence proportions of the subsets in $D_+$ are expected to be larger or equal to 0.5.
	
	Similar to HUMO, {\em r}-HUMO alternately selects the pairs in $D_-$ and $D_+$ for manual inspection to enforce precision and recall guarantees. Note that compared with first working on $D_-$ and then on $D_+$, working alternately on D- and D+ would result in the human-labeled pairs with a wider variety of machine metric values between iterations. Risk analysis based on the human-labeled results could therefore be less biased. In the rest of this subsection, we first describe the processes of pair selection on $D_-$ and $D_+$, and then present the algorithm to enforce quality guarantees based on them.
	
\vspace{0.04in}	
\hspace{-0.2in}{\bf Pair Selection in $D_-$.}	According to the monotonicity assumption of precision, the pairs in $D_i$ (the rightmost subset in $D_-$) have higher probabilities of being {\em equivalent} than any other subset in $D_-$. Accordingly, they are at the highest risk to be mislabeled by the machine. Therefore, {\em r}-HUMO sets the initial set of candidate pairs, denoted by $D_{-}'$, to be $D_i$, or $D_-'=D_i$. It then iteratively selects the pairs in $D_-'$ in a risk-wise decreasing order for human inspection. We define the Marginal Equivalence Proportion (MEP) of selection by
\begin{equation}
   MEP(D_-')=\frac{dM}{dN},
\end{equation}
in which the variables $N$ and $M$ represent the number of inspected pairs and the number of equivalent pairs among the inspected pairs respectively, and the differential operator ``\emph{d}'' represents the increment of the variables M and N in a period of human inspection. It can be observed that if risk estimation is effective, $MEP(D_-')$ would decrease as the selection proceeds. For simplicity of presentation, we denote the expected equivalence proportion of a subset $D_i$ by $EP(D_i)$. Iterative pair selection on $D_-'$ would stop once either of the two following conditions is satisfied:
\begin{enumerate}
    \item The expected equivalence proportion of the remaining pairs in $D_-'$ falls below the expected equivalence proportion of the rightmost uninspected unit subset adjacent to $D_-'$ in $D_-$. Denoting the rightmost unit subset by $D_j$, we can specify the condition by $EP(D_-')<EP(D_j)$;
	\item The marginal equivalence proportion of selection, $MEP(D_-')$, falls below the expected equivalence proportion of the remaining pairs in $D_-'$, or $MEP(D_-')<EP(D_-')$.
\end{enumerate}

It can be observed that if the first condition is triggered, it means that the pairs mislabeled by the machine can be more easily found in the unit subset $D_j$ instead of $D_-'$. If the second condition is triggered, it means that risk analysis on $D_-'$ has become ineffectual; all the remaining pairs in $D_-'$ should therefore be either automatically labeled by the machine or manually labeled by the human. In both cases, {\em r}-HUMO would merge the current candidate set and the rightmost uninspected unit subset adjacent to $D_-'$, $D_j$, to constitute a new candidate set, $D_-'=D_-'\cup D_j$. It would then re-estimate pair risk based on the updated human-labeled results and begin a new pair pick-out iteration on the new $D_-'$. To handle the case when pair selection stops too early, {\em r}-HUMO sets a threshold for the number of inspected pairs in each iteration. If the number of pairs chosen for inspection is less than the threshold number, it would assign all the remaining pairs in the rightmost unit subset in the candidate set $D_-'$ to the human.

\vspace{0.03in}	
\hspace{-0.2in}{\bf Pair Selection in $D_+$.}
  The process of human workload selection in $D_+$ is similar. We denote the candidate set considered for manual inspection in $D_+$ by $D_+'$. {\em r}-HUMO iteratively selects the pairs in $D_+'$ in a risk-wise decreasing order for human inspection. Since the pairs with lower similarities are at higher risk to be mislabeled by the machine, $D_+'$ is initially set to be the leftmost unit subset in $D_+$, or $D_+'=D_{i+1}$ in Fig.~\ref{fig:r-HUMO}. For simplicity of presentation, we denote the leftmost uninspected unit subset adjacent to $D_+'$ in $D_+$ by $D_k$. Iterative pair selection in $D_+'$ would stop once either of the two following conditions is satisfied:
\begin{enumerate}
    \item The expected equivalence proportion of the remaining pairs in $D_+'$ exceeds the expected equivalence proportion of the leftmost uninspected unit subset adjacent to $D_+'$ in $D_+$, or $EP(D_+')>EP(D_k)$;
	\item The marginal equivalence proportion of pair selection in $D_+'$ exceeds the expected equivalence proportion of the remaining pairs in $D_+'$, or $MEP(D_+')>EP(D_+')$.
\end{enumerate}
In both cases, {\em r}-HUMO would merge $D_+'$ and $D_k$ to constitute a new candidate set, $D_+'=D_+'\cup D_k$. It would then re-estimate pair risk based on the updated human-labeled results and begin a new iteration of pair pick-out on the new $D_+'$. Similar to the case of $D_-$, {\em r}-HUMO sets a lower threshold for the number of inspected pairs in each iteration.

\vspace{0.04in}	
\hspace{-0.2in}{\bf Algorithm.} The process of real-time human workload selection is sketched in Algorithm~\ref{alg:humanworkloadselection}. It alternately selects the pairs in $D_-$ and $D_+$ for manual inspection. More details on the algorithm can be found in our technical report~\cite{rhumotechrep}, but omitted here due to space limit.
	
\begin{algorithm}[t]
\caption{Real-time Human Workload Selection in {\em r}-HUMO.}
\label{alg:humanworkloadselection}
\While{$recall_L<\beta$ \bf{or} $precision_L<\alpha$}
{
    \If{$recall_L<\beta$}
    {
        Iteratively select pairs in $D_-'$ until one of the stop conditions is triggered\;
				Update $D_-'$\;
        Re-estimate pair risk\;
    }
    \If{$precision_L<\alpha$}
    {
        Iteratively select pair in $D_+'$ until one of the stop conditions is triggered\;
				Update $D_+'$\;
        Re-estimate pair risk\;
    }
}
\end{algorithm}

\subsection{Batch Human Workload Selection}

  In real-time human workload selection, given a set of candidate pairs, {\em r}-HUMO iteratively selects the riskiest pair for manual inspection, and updates in real time the equivalence proportion expectation of the remaining candidate pairs and marginal equivalence proportion of selection, based on the human label, to guide the next selection. In other words, it needs to wait for human labeling result until it can generate the next task for human inspection. This setting may be impractical in a real human-machine cooperation environment, as workers do not always respond in a real-time manner. Therefore, in this subsection, we propose a slightly batch version of {\em r}-HUMO, which allow human to inspect multiple pairs at each iteration.

    The idea of batch selection is to predict the least number of pairs that need to be manually inspected such that either of the two stop conditions can be satisfied. We take pair selection in $D_-$ as example. The case for $D_+$ is similar, thus omitted here. Consider the first condition $EP(D_-')<EP({D_j})$, which specifies that the expected equivalence proportion of the remaining pairs in the candidate set $D_-'$ falls below the expected equivalence proportion of the rightmost uninspected unit subset adjacent to $D_-'$ in $D_-$. With the assumption that the marginal equivalence proportion of selection in $D_-'$ decreases monotonously as the selection proceeds, the minimal equivalence proportion of the remaining pairs in $D_-'$ after $x$ pairs are selected for manual inspection can be represented by
\begin{equation}
   EP_{L}=\frac{EP(D_-')\cdot n-MEP(D_-')\cdot x}{n-x},
\end{equation}	
in which $n$ represents the total number of pairs in the original $D_-'$, and $EP(D_-')$ represents its expected equivalence proportion. As a result, the least number of pairs, which need to be manually inspected such that the first condition can be satisfied, can be represented by
\begin{equation}
    N_1=\frac{n\cdot (EP(D_-')-EP(D_j))}{MEP(D_-')-EP(D_j)}.
\end{equation}

  Now we consider the second condition, $MEP({D_-'})<EP(D_-')$. Suppose that the current value of the marginal equivalence proportion is denoted by $MEP(D_-')$, which is represented by $\frac{m'}{n'}$, and the current value of the equivalence proportion is denoted by $EP(D_-')$.
In the worst case that the following $x$ manually inspected pairs are all \emph{inequivalent} pairs, the minimal value of the latest marginal equivalence proportion can be represented by $\frac{m'}{n'+x}$; at the same time, the maximal value of the latest equivalence proportion can be represented by $\frac{EP(D_-')\cdot n}{n-x}$. Therefore, the least number of pairs, which need to be manually inspected for the second condition to be satisfied, can be represented by
\begin{equation}
  N_2=\frac{m'\cdot n-EP(D_-')\cdot n'\cdot n}{m'+EP(D_-')\cdot n}.
\end{equation}	
	
   In summary, at each interaction, r-HUMO can select at least min\{$N_1,N_2$\} pairs in $D_-'$ for manual inspection before either of the two stop conditions can be satisfied.

\subsection{Quality Assurance}

As in HUMO, the lower bounds of the achieved precision and recall levels of an {\em r}-HUMO solution are represented by Eq.~\ref{eq:precision-bound} and Eq.~\ref{eq:recall-bound} respectively. In this subsection, we present the GPR process to approximate the equivalence proportions of unit subsets and then describe how to compute the lower and upper bounds of the number of equivalent pairs in a given set based on GPR estimates.
	
\vspace{0.04in}
\hspace{-0.25in} {\bf GPR Approximation.}	As HUMO, {\em r}-HUMO samples a fraction of unit subsets to estimate their equivalence proportions and then uses GPR to approximate the equivalence proportions of other unit subsets. Suppose that $k$ unit subsets are sampled, their observed equivalence proportions are denoted by $\mathsf{R} = [\mathsf{R}_1, \mathsf{R}_2, \ldots, \mathsf{R}_k]^T$. For each unit subset, we also denote its average pair similarity by $v_i$. Accordingly, we denote the average pair similarities of the $k$ sampled subsets by $V = [v_1, v_2, \ldots, v_k]^T$.  Given a new unit subset $D_*$ and its average pair similarity value $v_*$, GPR assumes that the random variables of $[V^T, v_*]^T$ satisfy a joint Gaussian distribution, which is represented by
  \begin{equation}
  \begin{bmatrix} V \\ v_* \end{bmatrix} \sim
  \mathcal{N}\left(0, \begin{bmatrix} \mathbf{K}(V, V) & \mathbf{K}(V, v_*) \\ \mathbf{K}(v_*, V) & \mathbf{K}(v_*, v_*) \end{bmatrix}\right),
  \label{eq:jointdistribution}
  \end{equation}
in which $\mathbf{K}(\cdot ,\cdot)$ represents the covariance matrix. The details of how to compute the covariance matrix can be found in ~\cite{rasmussen2006gaussian}. Based on Eq.~\ref{eq:jointdistribution}, the distribution of the equivalence proportion of $S_*$, $R_*$, can be represented by the following Gaussian function
\begin{equation}
  \mathsf{R}_* \sim N\left(\bar{\mathsf{R}}_*, \sigma_{\mathsf{R}_*}^2\right),
\end{equation}
in which the mean of $\mathsf{R}_*$, $\bar{\mathsf{R}}_*$, can be represented by
\begin{equation}
  \bar{\mathsf{R}}_* = \mathbf{K}(v_*, V) \cdot \mathbf{K}^{-1}(V, V) \cdot \mathsf{R},
  \label{eq:gpr:mean}
\end{equation}
and the variance of $\mathsf{R}_*$,  $\sigma_{\mathsf{R}_*}^2$, can be represented by
\begin{equation}
  \sigma_{\mathsf{R}_*}^2 = \mathbf{K}(v_*, v_*) - \mathbf{K}(v_*, V) \cdot \mathbf{K}^{-1}(V, V) \cdot \mathbf{K}(V, v_*).
\label{eq:gpr:variance}
\end{equation}

\vspace{0.02in}
\hspace{-0.25in} {\bf Bound Estimation.}
	Provided with the Gaussian distributions of the equivalence proportions of unit subsets, the number of equivalent pairs in any given set consisting of multiple unit subsets can be estimated by aggregating the distributions of unit subsets. Suppose that the pair set of $D_*$ consists of $t$ unit subsets, $D_* = D_*^1\cup D_*^2\cup \ldots D_*^t$, the total number of pairs in $D_*^i$ ($1\leq i\leq t$) is denoted by $n_*^i$, and the average similarity value of the pairs in $D_*^i$ by $v_*^i$.  Then, the total number of equivalent pairs in $D_*$, denoted by $m_*$, can be represented by
\begin{equation}
  \mathsf{m}_* \sim N\left(\bar{\mathsf{m}}_*, \sigma_{\mathsf{m}_*}^2\right),
\end{equation}
in which the mean, $\bar{\mathsf{m}}_*$, can be represented by
\begin{equation}
  \bar{m}_* = \sum_{i=1}^{t}n_*^i\cdot\bar{\mathsf{R}}_*^i,
\label{eq:aggregation-mu}
\end{equation}
and the variance, $\sigma_{\mathsf{m}_*}^2$, can be represented by
\begin{equation}
\begin{array}{l}
\sigma_{\mathsf{m}_*}^2 = \sum\limits_{1 \le i \le t,1 \le j \le t} {n_*^i}  \cdot n_*^j \cdot cov(v_*^i,v_*^j) ,
\end{array}
\label{eq:aggregation-sigma}
\end{equation}
in which $cov(v_*^i, v_*^j)$ is the covariance between the two estimates.

  In {\em r}-HUMO, some pairs in a unit subset may be inspected by the human while others are automatically labeled by the machine. In other words, $D_H$, $D_-$ and $D_+$ may contain a fraction of the pairs in a unit subset. Consider a pair set consisting of
$t$ subsets, ${D_*}' = {D_*^1}'\cup {D_*^2}'\cup \ldots {D_*^t}'$, in which ${D_*^i}'$ denotes the set of remaining pairs in $D_*^i$ with some of the pairs selected for manual inspection and ${D_*^i}'\neq \emptyset$. Suppose that there are $r$ equivalent pairs among all the pairs inspected by the human in $D_*$, or the pairs in $D_*-{D_*}'$. On the number of equivalent pairs in ${D_*}'$, we have Lemma~\ref{le:remaining}, whose proof is straightforward, thus omitted here due to space limit.

\begin{lemma}
Provided that the number of equivalent pairs in $D_*$ can be represented by the Gaussian function of $N(\bar{m}_*, \sigma_{m_*}^2)$, the number of equivalent pairs in $D_*'$, $m_*'$, can thus be represented by the Gaussian function of
\begin{equation}
m_*' \sim N(\bar{m}_*-r, \sigma_{m_*}^2).
\label{eq:remaining}
\end{equation}
\label{le:remaining}
\end{lemma}

Note that the correctness of Lemma~\ref{le:remaining} depends on the non-emptiness of unit subsets ${D_*^i}'$. If all the pairs of a unit subset have been chosen for manual inspection and it becomes empty, its estimation covariance with any other estimate on other unit subsets would become zero.

   Finally, given the confidence level of $\theta$, the lower and upper bounds of the number of equivalent pairs in a subset $D_*$ can be represented by
\begin{equation}
  [\bar{m}_* - \mathcal{Z}_{(1-\theta)} \cdot \sigma_{m_*}, \bar{m}_* + \mathcal{Z}_{(1-\theta)} \cdot \sigma_{m_*}],
\label{eq:lbub}
\end{equation}
in which $\mathcal{Z}_{(1-\theta)}$ is the $(1-\frac{1-\theta}{2})$ point of {\em standard normal distribution}.

%% file: riskanalysis.tex
\section{Risk Analysis} \label{sec:riskmodel}

   Risk analysis of {\em r}-HUMO is performed on the human-labeled results. Given a candidate pair set, {\em r}-HUMO iteratively selects the most risky pairs in it for manual inspection. It can be said that the performance of {\em r}-HUMO depends on the effectiveness of risk analysis. Motivated by its success in modern portfolio investment theory~\cite{Markowitz1991Foundations, Rockafellar2002Conditional, Acerbi2002Expected}, {\em r}-HUMO employs the metric of Conditional Value at Risk (CVaR) to measure the risk of pairs being mislabeled by the machine.
	
	In the portfolio risk theory, given the confidence level of $\theta$, CVaR is defined, in a conservative way, to be the expected loss incurred in the $1-\theta$ worst case. Formally, given the loss function $z(X) \in {L^p}(F)$ of a portfolio $X$ and the confidence level of $\theta$, the metric of CVaR is defined as
\begin{equation}
CVa{R_\theta }(X) = \frac{1}{{1 - \theta }}\int\limits_0^{1 - \theta } {Va{R_{1 - \gamma }}(X)} d\gamma,
\end{equation}
where $Va{R_{1 - \gamma }}(X)$ represents the minimum loss incurred in the $\gamma$ worst case and can be formally represented by
\begin{equation}
Va{R_{1 - \gamma }}(X) = \inf \{ {z_*}:P(z(X) \ge {z_*}) \le \gamma \}.
\end{equation}

  According to CVaR's definition, risk measurement requires the labeled pair's potential loss estimation. Intuitively, this loss refers to the probability of the pair's label being incorrect. As typical in CVaR evaluation, {\em r}-HUMO represents the equivalence probability of a pair by a Gaussian distribution and estimates the potential loss based on it. It considers a pair as a portfolio consisting of multiple stocks. Each stock corresponds to a feature of the pair and its loss corresponds to its corresponding feature's equivalence or inequivalence probability. In the rest of this section, we first describe how to extract features from human-labeled pairs, and then present the metric of risk measurement and analyze its complexity.

\subsection{Feature Extraction}

   For general purpose, the features used for risk analysis should have the following three desirable properties: (1) they could be easily extracted from the human-labeled pairs; (2) they should be evidential, or indicative of the equivalence status of a pair; (3) they should be to a large extent independent of the machine metric used in ordering pairs in the first place. It can be observed that in {\em r}-HUMO, the pairs are generally chosen into the candidate set in the order dictated by a pre-specified machine metric. Therefore, the features independent of the machine metric would be more effective than the non-independent ones in differentiating the pairs with similar metric values in terms of mislabeling risk.

	{\em r}-HUMO extracts two types of features from the human-labeled pairs, $Same(t_i)$ and $Diff(t_i)$, in which $t_i$ represents a token, $Same(t_i)$ means that $t_i$ appears in both records in a pair, and $Diff(t_i)$ means that $t_i$ appears in one and only one record in a pair. It can be observed that these two features are evidential and easily extractable. Moreover, they were not used in the existing classification metrics for ER. Our risk model assumes that the equivalence probability of a feature satisfies a normal distribution. Given a feature $f$ and a set of human-labeled pairs with the feature $f$, $D_f$, the expectation of the equivalence probability of $f$ can be represented by
\begin{equation}
E(f) = \frac{{|D_{f+}|}}{{|{D_f}|}},
\end{equation}
in which $D_{f+}$ denotes the set of equivalent pairs in $D_f$. Its variance can also be represented by
\begin{equation}
V(f) = \frac{1}{{|{D_f}| - 1}}\sum\limits_{{d_i} \in {D_f}} {{{(L({d_i}) - E(f))}^2}},
\end{equation}
in which $L(d_i)$ denotes the manual label of a pair $d_i$ in $D_f$, $L(d_i)=1$ if $d_i$ is labeled as \emph{matching} and $L(d_i)=0$ if $d_i$ is labeled as \emph{unmatching}.

\subsection{Risk Measurement}

    In this subsection, we first propose the risk model for the case that features are independent; we then describe how to handle the more complicated case where the features are not independent.
		
	According to the theory of portfolio investment, a pair's equivalence probability distribution can be represented by the weighted linear combination of the distributions of its features. Therefore, provided with the Gaussian distributions of features, the equivalence probability expectation of a pair $d$, can be represented by
\begin{equation}
E(d) = \sum\limits_{{f_i} \in {F_d}} {{w}_d({f_i}) \cdot E({f_i})},
\label{eq:pair-mean}
\end{equation}
in which ${F_d}$ denotes the set of features contained in $d$, and ${w}_d({f_i})$ denotes the weight of $f_i$ in $d$. Its variance can also be represented by
\begin{equation}
V(d) = \sum\limits_{{f_i} \in {F_d}} {w_d({f_i})^2 \cdot V({f_i})}.
\label{eq:pair-variance}
\end{equation}
The weight of the feature ${f_i}$ in $d$ is defined as
\begin{equation}
{w}_d({f_i}) = \frac{{w({f_i})}}{{\sum\limits_{{f_j} \in {F_d}} {w({f_j})} }},
\label{eq:featureweight}
\end{equation}
where $w({f_i})$ denotes the absolute weight of the feature $f_i$.

  In Eq.~\ref{eq:featureweight}, the absolute feature weights can be simply set to 1: each feature is equally powerful in predicting a given pair's equivalence probability. In most practical scenarios, this assumption may not hold true. Therefore, {\em r}-HUMO uses the concept of information value~\cite{hababou2006variable, Zhang2017Three} to determine feature weight. The details on setting feature weight can be found in our technical report~\cite{rhumotechrep}, but omitted here due to space limit.

   Given a pair $d$, we denote its equivalence probability by $x$, and the probability density function and cumulative distribution function of $x$ by ${pd{f_d}}(x)$ and ${cd{f_d}}(x)$ respectively. Suppose that $d$ is originally labeled by the machine as {\em unmatching}. Then, the probability of $p$ being mislabeled by the machine is equal to $x$. Accordingly, the worst-case loss of $d$ corresponds to the case that $x$ is maximal. Therefore, given the confidence level of $\theta$, the CVaR of $d$ is the expectation of $z=x$ in the $1-\theta$ cases where $x$ is from $cd{f_d}^{ - 1}(\theta)$ to $+ \infty$. Formally, the CVaR risk of a pair $d$ with the machine label of {\em unmatching} can be estimated by
\begin{equation}
CVa{R_\theta }(d) =
{\frac{1}{{1 - \theta }}\int\limits_{cd{f_d}^{ - 1}(\theta )}^{ + \infty } {pd{f_d}} (x) \cdot xdx}.
\end{equation}

Otherwise, $d$ is originally labeled by machine as {\em matching}. Then the potential loss of $d$ being mislabeled by machine is equal to $1-x$. Therefore, the CVaR risk of a pair $d$ with the machine label of {\em matching} can be estimated by
\begin{equation}
CVa{R_\theta }(d) =
{\frac{1}{{1 - \theta }}\int\limits_{ - \infty }^{cd{f_d}^{ - 1}(1 - \theta )} {pd{f_d}} (x) \cdot (1 - x)dx}.
\end{equation}

  The above-described process assumes that the extracted features are independent. Unfortunately, this assumption may not hold true in practice. In the case that the features contained by a pair $d$ are {\em not} independent, {\em r}-HUMO again borrows the idea of modern portfolio investment theory and introduces the covariances between features into the process of risk measurement. The technical details on how to handle feature dependency can be found in our technical report~\cite{rhumotechrep}, but omitted here due to space limit.

\subsection{Complexity Analysis}

   It can be observed that in each iteration, the total frequency of feature occurrence is bound by {\bf\em O}($l\cdot n$), in which $n$ denotes the total number of pairs in a workload and $l$ denotes the maximal number of tokens in a pair. As a result, the time complexity of computing the feature distributions in each iteration is bounded by {\bf\em O}($m\cdot n$). Accordingly, without considering feature dependency, the time complexity of computing the CVaR risk of the candidate pairs in each iteration is also bounded by {\bf\em O}($l\cdot n$). Since the number of iterations is at most {\bf\em O}($n$), the total time complexity of risk analysis can be represented by {\bf\em O}($l\cdot n^2$).
   Therefore, we have Theorem.~\ref{le:remaining}, whose proof follows naturally from the above analysis.
\begin{theorem}
    The space and time complexities of risk analysis without considering feature dependency can be represented by {\bf\em O}($l\cdot n$) and {\bf\em O}($l\cdot n^2$) respectively, in which $n$ denotes the total number of pairs in a workload and $l$ denotes the maximal number of tokens in a pair.
\label{le:remaining}
\end{theorem}

%% file: experiment.tex
\section{Experimental Study} \label{sec:Experiments}

  In this section, we empirically evaluate the performance of {\em r}-HUMO on real data by comparative study. We compare {\em r}-HUMO with the state-of-the-art alternative HUMO \cite{humodemo, humo}, which can enforce both precision and recall, as well as two baselines. Note that most of existing ER techniques can not enforce the quality guarantees measured by precision and recall. Their comparative performance evaluation is therefore beyond the scope of this paper. However, we also compare r-HUMO with the active learning-based approach (denoted by ACTL) \cite{arasu2010active}, which can at least enforce precision. ACTL can maximize recall while ensuring a pre-specified precision level. It estimates the achieved precision level of a labeling solution by sampling. As a result, ACTL also requires manual inspection. We compare {\em r}-HUMO with ACTL on the achieved quality and the required manual cost.
	
	The rest of this section is organized as follows: Subsection~\ref{sec:setup} describes the experimental setup. Subsection~\ref{sec:achievedquality} evaluates quality guarantee of {\em r}-HUMO. Subsection~\ref{sec:ihumovshumo} compares {\em r}-HUMO with HUMO. Subsection~\ref{sec:baseline} compares {\em r}-HUMO with two baselines. Subsection~\ref{sec:ihumovsactl} compares {\em r}-HUMO with ACTL. Subsection~\ref{sec:batch} evaluates the efficacy of batch human workload selection.  Finally, Subsection~\ref{sec:scalability} evaluates the efficiency and scalability of {\em r}-HUMO.

\subsection{Experimental Setup} \label{sec:setup}

Our evaluation is conducted on the three real datasets, whose details are described as follows:
\begin{itemize}
\item DBLP-Scholar\footnote{available at https://dbs.uni-leipzig.de/file/DBLP-Scholar.zip} (denoted by DS): The DS dataset contains 2616 publication entities from DBLP and 64263 publication entities from Google Scholar. The experiments match the DBLP entries with the Scholar entries.
\item Abt-Buy\footnote{available at https://dbs.uni-leipzig.de/file/Abt-Buy.zip} (denoted by AB): The AB dataset contains 1081 product entities from Abt.com and 1092 product entities from Buy.com. The experiments match the Abt entries with the Buy entries.
\item Songs\footnote{available at http://pages.cs.wisc.edu/\~{}anhai/data/falcon\_data/songs} (denoted by SG): The SG dataset contains 1000000 song entities, some of which refer to the same songs. The experiments match the song entries in the same table.
\end{itemize}

  Our empirical study uses pair similarity as the machine metric. It computes pair similarity by aggregating attribute similarities with weights \cite{humodemo, humo}. Specifically, on the DS dataset, Jaccard similarity of the attributes {\em title} and \emph{authors}, and Jaro-Winkler distance of the attribute {\em venue} are used; on the AB dataset, Jaccard similarity of the attributes \emph{product name} and \emph{product description} are used; on the SG dataset, Jaccard similarity of the attributes \emph{song title}, \emph{release information} and \emph{artist name}, Jaro-Winkler distance of the attributes \emph{song title} and \emph{release information}, and number similarity of the attributes \emph{duration}, \emph{artist familiarity}, \emph{artist hotness} and \emph{year} are used. The weight of each attribute is determined by the number of its distinct values. As in \cite{arasu2010active, humodemo, humo}, we use the blocking technique to filter the instance pairs unlikely to match. Specifically, the DS workload contains the instance pairs whose aggregated similarity values are no less than 0.2. Similarly, the aggregated similarity value thresholds for the AB and SG workloads are set to 0.05 and 0.2. After blocking, the DS workload has 100077 pairs, 5267 among them are equivalent pairs; the AB workload has 313040 pairs, 1085 among them are equivalent pairs; the SG workload has 289893 pairs and 13756 among them are equivalent pairs.

\begin{figure}[h!]
\centering
\includegraphics[width=0.32\linewidth]{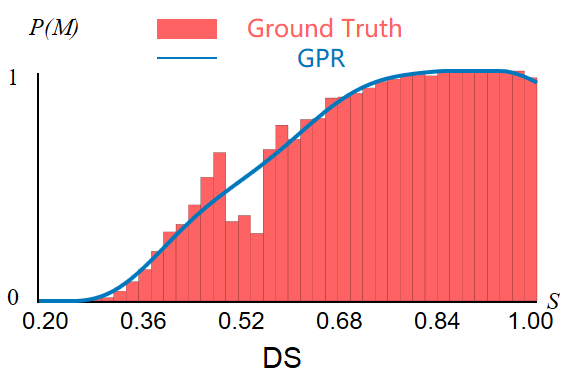}
\includegraphics[width=0.32\linewidth]{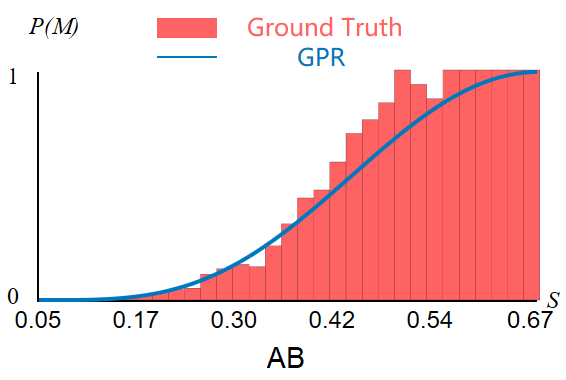}
\includegraphics[width=0.33\linewidth]{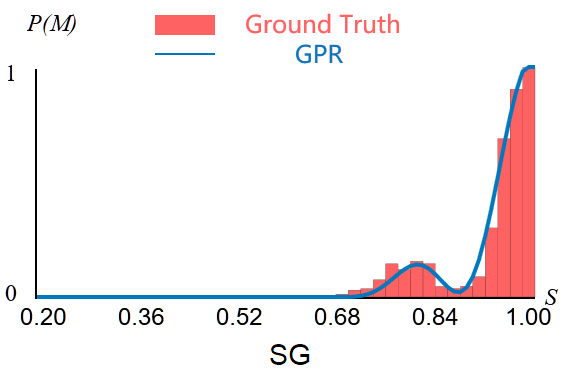}
\caption{the Equivalence Proportions of Unit Subsets with regard to Pair Similarity.}
\label{fig:matchdistribution}
\end{figure}

  As in the HUMO implementation, our {\em r}-HUMO implementation partitions an ER workload into disjoint unit subsets, each of which contains
the same number of instance pairs. The number of instance pairs contained by each subset is set to 200. The equivalence proportions of the uit subsets with regard to pair similarity on the three workloads are presented in Figure.~\ref{fig:matchdistribution}. It can be observed that on all the three test datasets, monotonicity of precision is a general trend. Specifically, on the SG dataset, which to the largest extent violates the assumption among them, monotonicity is not satisfied only in a small range of similarity value, between 0.70 and 0.82.

   To balance the sampling cost and the accuracy of equivalence proportion approximation, as in the HUMO implementation \cite{humodemo, humo}, {\em r}-HUMO sets both lower and upper limits on sampling cost, which is measured by the proportion of sampled unit subsets among all the subsets. In our experiments, the range of the sampling proportion is set between 3\% and 5\%. We observe that considering feature dependency in risk analysis for {\em r}-HUMO can only marginally improve the performance on the test workloads. We therefore report the results of {\em r}-HUMO without considering feature dependency. The evaluation results of the {\em r}-HUMO implementation considering feature dependency can instead be found in our technical report \cite{rhumotechrep}.

   Note that in {\em r}-HUMO, different runs may generate different labeling solutions due to sampling randomness. For each experiment, we therefore run the program 20 times on each workload and report the averaged result. In our experiments, we have the ground-truth labels of all the test pairs. The ground-truth labels are originally hidden; whenever manual inspection is called, they are provided to the program.

\subsection{Quality Enforcement} \label{sec:achievedquality}

   In the experiments, we specify 7 scales of quality requirement, whose precision and recall are set at
different levels at 0.8, 0.825, 0.85, 0.875, 0.9, 0.925 and 0.95 respectively. The confidence level on quality guarantee is set at 0.9.

\begin{table}[!h]
\centering
\caption{Evaluation of Quality Enforcement.}
\vspace{-0.05in}
\label{tb:quality}
\setlength{\tabcolsep}{2.2mm}{
\begin{tabular}{|c|c|ccc|}
\hline
\multirow{2}{*}{\begin{tabular}[c]{c}Dataset\end{tabular}} &Required Quality& \multicolumn{3}{c|}{Achieved Quality} \\ \cline{2-5}
&$\alpha$=$\beta$     & $\alpha$ & $\beta$      & SR(\%)     \\ \hline
\multirow{6}{*}{\begin{tabular}[c]{c}DS\end{tabular}}
&0.825    & 0.9079 & 0.8459 & 100    \\
&0.850    & 0.9098 & 0.8657 & 100    \\
&0.875    & 0.9124 & 0.8904 & 100    \\
&0.900    & 0.9248 & 0.9150 & 100    \\
&0.925    & 0.9497 & 0.9391 & 100    \\
&0.950    & 0.9748 & 0.9628 & 95     \\ \hline
\multirow{7}{*}{\begin{tabular}[c]{c}AB\end{tabular}}
&0.800    & 0.9629 & 0.8546 & 100    \\
&0.825    & 0.9630 & 0.8589 & 100    \\
&0.850    & 0.9635 & 0.8718 & 100    \\
&0.875    & 0.9643 & 0.8920 & 100    \\
&0.900    & 0.9651 & 0.9112 & 100    \\
&0.925    & 0.9671 & 0.9398 & 100    \\
&0.950    & 0.9693 & 0.9508 & 90     \\ \hline
\multirow{7}{*}{\begin{tabular}[c]{c}SG\end{tabular}}
&0.800    & 0.9663 & 0.8957 & 90     \\
&0.825    & 0.9671 & 0.9201 & 90     \\
&0.850    & 0.9678 & 0.9417 & 90     \\
&0.875    & 0.9683 & 0.9581 & 90     \\
&0.900    & 0.9686 & 0.9660 & 90     \\
&0.925    & 0.9687 & 0.9694 & 90     \\
&0.950    & 0.9746 & 0.9708 & 90     \\ \hline
\end{tabular}}
\end{table}
	
	The detailed evaluation results are presented in Table~\ref{tb:quality}, in which $\alpha$ denotes precision, $\beta$ denotes recall, and $SR$ denotes success rate (the percentage of the successful runs achieving required quality levels among all the runs). On DS, the initial machine labeling solution achieves precision and recall levels above 0.8; we therefore do not report the results in the table. It can be observed that {\em r}-HUMO is effective in enforcing quality guarantees. On all the three workloads, the achieved quality levels are considerably above the required levels in most cases and the achieved success rates consistently exceed the confidence level of 0.9.

\subsection{{\em r}-HUMO vs HUMO} \label{sec:ihumovshumo}

  We compare the manual cost consumed by {\em r}-HUMO and HUMO given the same quality requirement. Bear in mind that the manual cost includes both the sampling cost and the cost of manually inspecting the pairs in $D_H$. Since {\em r}-HUMO and HUMO consume the same amount of sampling cost in each run, we compare the size of $D_H$ (excluding sampled pairs). On all the three datasets, the consumed sampling cost is very close to the upper limit of 5\% in most runs; averagely, the sampling cost of DS, AB and SG are 4.93\%, 4.94\% and 4.96\% respectively.
	
	The comparative results of {\em r}-HUMO and HUMO are presented in Table~\ref{tb:result}. It can be observed that given the same quality requirement, {\em r}-HUMO consistently consumes less human cost than HUMO and the cost difference between them is considerable in most cases. {\em r}-HUMO iteratively selects a few mislabeled pairs from a set of mostly correctly labeled pairs. With the stricter quality requirement, the task becomes more challenging because it has to identify the mislabeled pairs in the unit subsets with increasingly low mislabeling proportion. In other words, in {\em r}-HUMO, the required human cost for finding out a fixed number of mislabeled pairs would increase with the decreasing equivalence proportion. Accordingly, its advantage over HUMO would gradually become smaller as pair selection proceeds. As a result, as shown in Table~\ref{tb:result}, the performance difference between {\em r}-HUMO and HUMO narrows down as quality requirement is enhanced.

\begin{table}[!h]
\centering
\caption{Performance Comparison between {\em r}-HUMO and HUMO.}
\vspace{-0.05in}
\label{tb:result}
\setlength{\tabcolsep}{1.4mm}{
\begin{tabular}{|c|c|cc|c|}
\hline
\multirow{2}{*}{\begin{tabular}[c]{c}Data\\set\end{tabular}} &Required Quality & \multicolumn{3}{c|}{Size of $D_H$ (excl. samples)}  \\ \cline{2-5}
&$\alpha$=$\beta$ & HUMO & {\em r}-HUMO & Reduction(\%)\\ \hline
\multirow{6}{*}{\begin{tabular}[c]{c}DS\end{tabular}}
&0.825   & 108         & \textbf{37}      & 65.74                \\
&0.850   & 463         & \textbf{151}     & 67.39                \\
&0.875   & 927         & \textbf{302}     & 67.42                \\
&0.900   & 1255        & \textbf{538}     & 57.13                \\
&0.925   & 1852        & \textbf{967}     & 47.79                \\
&0.950   & 2802        & \textbf{1786}    & 36.26                \\ \hline
\multirow{7}{*}{\begin{tabular}[c]{c}AB\end{tabular}}
&0.800   & 4628        & \textbf{4144}    & 10.46                \\
&0.825   & 6056        & \textbf{5664}    & 6.47                 \\
&0.850   & 7894        & \textbf{7179}    & 9.06                 \\
&0.875   & 10239       & \textbf{8328}    & 18.66                \\
&0.900   & 13495       & \textbf{10662}   & 20.99                \\
&0.925   & 28273       & \textbf{27380}   & 3.16                 \\
&0.950   & 34649       & \textbf{34136}   & 1.48                 \\ \hline
\multirow{7}{*}{\begin{tabular}[c]{c}SG\end{tabular}}
&0.800   & 62940       & \textbf{28007}   & 55.50                \\
&0.825   & 66672       & \textbf{34615}   & 48.08                \\
&0.850   & 68598       & \textbf{43055}   & 37.24                \\
&0.875   & 71667       & \textbf{59654}   & 16.76                \\
&0.900   & 75698       & \textbf{74156}   & 2.04                 \\
&0.925   & 83156       & \textbf{81903}   & 1.51                 \\
&0.950   & 89742       & \textbf{88190}   & 1.73                 \\ \hline
\end{tabular}}
\end{table}

   The results reported in Table~\ref{tb:result} are based on the GPR approximation that uses the samples, which consists of only a small portion (less than 5\%) of all the pairs in a workload. However, the results of GPR approximation may not be accurate under many circumstances. The estimation accuracy of GPR approximation may significantly affect the computation of $D_H$'s boundaries. To further validate the effectiveness of {\em r}-HUMO's pair prioritization strategy, we also execute {\em r}-HUMO and HUMO with the extreme assumption that the estimations of subset equivalence proportions are exactly right. In the new experimental setting, both {\em r}-HUMO and HUMO are given the exact number of equivalent pairs in each unit subset (instead of GPR estimation) beforehand as well as the same quality requirement. The comparative results on the three workloads are presented in Table~\ref{tb:truthgpr}. It can be observed that {\em r}-HUMO consistently outperforms HUMO in all the test cases, and compared with the previous setting of GPR approximation, {\em r}-HUMO outperforms HUMO by more considerable margins. \emph{Even though the results reported in Table~\ref{tb:truthgpr} are unrealistic in practical scenarios, they do illustrate the efficacy of r-HUMO's pair prioritization strategy and demonstrate that the performance advantage of r-HUMO over HUMO can increase with the tightness of GPR approximation.}
	
	Note that the bounds of precision and recall are estimated by aggregating the GPR approximations over all the unit subsets. Accordingly, small variations on the equivalence proportions of unit subsets can lead to comparatively large variations on the estimated precision and recall. Pair selection based on GPR approximation is therefore usually very conservative, resulting in human labeling cost much more than what is necessary for quality guarantee. In Table~\ref{tb:quality}, we can observe that most precision and recall achieved by r-HUMO exceed the required levels by considerable margins. If provided with ground-truth equivalence proportions on all the unit subsets, precision and recall could be estimated with certainty. The achieved precision and recall level of r-HUMO would therefore be much closer to the required levels. Since the efficacy of risk analysis decreases with the required quality, the cost reductions reported in Table~\ref{tb:truthgpr} are more considerable than those reported in Table~\ref{tb:result}. However, it does not mean that GPR approximation is ineffective. The accuracy of the bounds established by GPR approximation is necessary for quality guarantee. As shown in Figure~\ref{fig:matchdistribution}, the GPR approximation is generally accurate on all the three test datasets. The success of r-HUMO in quality guarantee, as shown in Table~\ref{tb:result}, also validates the effectiveness of the GPR approximation.

\begin{table}[!h]
\centering
\caption{{\em r}-HUMO vs HUMO with Ground-Truth Equivalence Proportions.}
\vspace{-0.05in}
\label{tb:truthgpr}
\setlength{\tabcolsep}{1.4mm}{
\begin{tabular}{|c|c|cc|c|}
\hline
\multirow{2}{*}{\begin{tabular}[c]{c}Data\\set\end{tabular}} &Required Quality & \multicolumn{3}{c|}{Size of $D_H$ (excl. samples)}     \\ \cline{2-5}
&$\alpha$=$\beta$ & HUMO & {\em r}-HUMO & Reduction(\%)\\ \hline
\multirow{5}{*}{\begin{tabular}[c]{c}DS\end{tabular}}
&0.850    & 443         & \textbf{78}        & 82.39                      \\
&0.875    & 949         & \textbf{221}       & 76.71                      \\
&0.900    & 1080        & \textbf{373}       & 65.46                      \\
&0.925    & 1574        & \textbf{586}       & 62.77                      \\
&0.950    & 2545        & \textbf{1020}      & 59.92                      \\ \hline
\multirow{7}{*}{\begin{tabular}[c]{c}AB\end{tabular}}
&0.800    & 2519        & \textbf{953}       & 62.17                      \\
&0.825    & 3997        & \textbf{1359}      & 66.00                      \\
&0.850    & 5971        & \textbf{2041}      & 65.82                      \\
&0.875    & 9617        & \textbf{2899}      & 69.86                      \\
&0.900    & 12671       & \textbf{4045}      & 68.08                      \\
&0.925    & 19466       & \textbf{14628}     & 24.85                      \\
&0.950    & 33226       & \textbf{32020}     & 3.77                       \\ \hline
\multirow{7}{*}{\begin{tabular}[c]{c}SG\end{tabular}}
&0.800    & 62298       & \textbf{12796}     & 79.46                      \\
&0.825    & 65101       & \textbf{14052}     & 78.42                      \\
&0.850    & 67983       & \textbf{15505}     & 77.19                      \\
&0.875    & 71006       & \textbf{17558}     & 75.27                      \\
&0.900    & 72820       & \textbf{19584}     & 73.11                      \\
&0.925    & 75640       & \textbf{23693}     & 68.68                      \\
&0.950    & 80869       & \textbf{30287}     & 62.55                      \\ \hline
\end{tabular}}
\end{table}

\vspace{-0.1in}

\subsection{{\em r}-HUMO vs Baselines} \label{sec:baseline}

 To further validate the efficacy of the proposed risk analysis technique, we also compare its performance with two baseline alternatives for pair selection, the native random strategy (denoted by Rand) and the simple strategy based on the distance from the center of the scale of a workload (denoted by CoS). In the experiments, the center of the scale is computed based on attribute similarities. Given the same GPR approximation result, we measure the achieved precision and recall of different strategies with the same amount of human cost budget. We set the cost budget as the number of manually inspected pairs required by {\em r}-HUMO to enforce the specified quality guarantees. The detailed comparative results are presented in Table \ref{tb:baseline}. It can be observed that given the same cost budget, {\em r}-HUMO achieves considerably better quality than both Rand and CoS. These experimental results show that {\em r}-HUMO is considerably more accurate in picking out the mislabeled pairs than Rand and CoS. They validate the efficacy of the proposed risk analysis technique.

\begin{table}[!h]
\centering
\caption{{\em r}-HUMO vs Two Baselines.}
\vspace{-0.05in}
\label{tb:baseline}
\setlength{\tabcolsep}{1.4mm}{
\begin{tabular}{|c|c|cc|cc|cc|}
\hline
\multirow{2}{*}{\begin{tabular}[c]{c}Dataset\end{tabular}} &\multirow{2}{*}{\begin{tabular}[c]{c}Cost\end{tabular}}
&\multicolumn{2}{c|}{{\em r}-HUMO} &\multicolumn{2}{c|}{CoS} &\multicolumn{2}{c|}{Rand} \\ \cline{3-8}
&& $\alpha$ & $\beta$  & $\alpha$ & $\beta$  & $\alpha$ & $\beta$  \\ \hline
\multirow{6}{*}{\begin{tabular}[c]{c}DS\end{tabular}}
&37      & 0.9079 & 0.8459 & 0.9075 & 0.8407 & 0.9075 & 0.8407     \\
&151     & 0.9098 & 0.8657 & 0.9075 & 0.8407 & 0.9075 & 0.8407     \\
&302     & 0.9124 & 0.8904 & 0.9076 & 0.8409 & 0.9075 & 0.8408     \\
&538     & 0.9248 & 0.9150 & 0.9076 & 0.8409 & 0.9075 & 0.8408     \\
&967     & 0.9497 & 0.9391 & 0.9076 & 0.8409 & 0.9075 & 0.8408     \\
&1786    & 0.9748 & 0.9628 & 0.9076 & 0.8409 & 0.9076 & 0.8409     \\ \hline
\multirow{7}{*}{\begin{tabular}[c]{c}AB\end{tabular}}
&4144    & 0.9629 & 0.8546 & 0.9475 & 0.6571 & 0.9475 & 0.6574     \\
&5644    & 0.9630 & 0.8589 & 0.9475 & 0.6571 & 0.9475 & 0.6576     \\
&7179    & 0.9635 & 0.8718 & 0.9475 & 0.6571 & 0.9475 & 0.6576     \\
&8328    & 0.9643 & 0.8920 & 0.9475 & 0.6571 & 0.9475 & 0.6576     \\
&10662   & 0.9651 & 0.9112 & 0.9475 & 0.6571 & 0.9475 & 0.6578     \\
&27380   & 0.9671 & 0.9398 & 0.9476 & 0.6589 & 0.9476 & 0.6587     \\
&34136   & 0.9693 & 0.9508 & 0.9476 & 0.6589 & 0.9477 & 0.6594     \\ \hline
\multirow{7}{*}{\begin{tabular}[c]{c}SG\end{tabular}}
&28007   & 0.9663 & 0.8957 & 0.9177 & 0.3381 & 0.9179 & 0.3390     \\
&34615   & 0.9671 & 0.9201 & 0.9177 & 0.3381 & 0.9180 & 0.3394     \\
&43055   & 0.9678 & 0.9417 & 0.9177 & 0.3382 & 0.9181 & 0.3399     \\
&59654   & 0.9683 & 0.9581 & 0.9179 & 0.3387 & 0.9184 & 0.3412     \\
&74156   & 0.9686 & 0.9660 & 0.9180 & 0.3393 & 0.9188 & 0.3429     \\
&81903   & 0.9687 & 0.9694 & 0.9180 & 0.3394 & 0.9190 & 0.3442     \\
&88190   & 0.9746 & 0.9708 & 0.9181 & 0.3399 & 0.9193 & 0.3454     \\ \hline
\end{tabular}}
\end{table}

\vspace{-0.1in}

\subsection{{\em r}-HUMO vs ACTL} \label{sec:ihumovsactl}

\begin{table}[!h]
\centering
\caption{{\em r}-HUMO vs ACTL on Recall given the Same Precision.}
\vspace{-0.05in}
\label{tab:compare-act}
\setlength{\tabcolsep}{0.8mm}{
\begin{tabular}{|c|c|c|c|c|c|c|}
\hline
\multirow{2}{*}{\begin{tabular}[c]{c}Data\\set\end{tabular}} &\multirow{2}{*}{\begin{tabular}[c]{c}Required\\Precision\end{tabular}} & \multicolumn{2}{c|}{Achieved Recall} & \multicolumn{2}{c|}{$\psi (\%)$} & \multirow{2}{*}{$\frac{\Delta \psi}{100 \cdot \Delta Recall}$}\\
\cline{3-6}
& & {\em r}-HUMO & ACTL & {\em r}-HUMO & ACTL & \\
\hline
\multirow{6}{*}{\begin{tabular}[c]{c}DS\end{tabular}}
&0.825 & 0.8459 & 0.8176 & 4.97  & 3.46 & 0.5335  \\
&0.850 & 0.8657 & 0.7999 & 5.08  & 3.70 & 0.2097  \\
&0.875 & 0.8904 & 0.8000 & 5.23  & 2.98 & 0.2489  \\
&0.900 & 0.9150 & 0.7662 & 5.47  & 3.17 & 0.1546  \\
&0.925 & 0.9391 & 0.7557 & 5.90  & 3.83 & 0.1129  \\
&0.950 & 0.9628 & 0.7273 & 6.71  & 2.56 & 0.1762  \\ \hline
\multirow{7}{*}{\begin{tabular}[c]{c}AB\end{tabular}}
&0.800 & 0.8546 & 0.1558 & 6.26  & 0.29 & 0.0854  \\
&0.825 & 0.8589 & 0.1395 & 6.75  & 0.28 & 0.0899  \\
&0.850 & 0.8718 & 0.1578 & 7.23  & 0.28 & 0.0973  \\
&0.875 & 0.8920 & 0.1152 & 7.60  & 0.27 & 0.0944  \\
&0.900 & 0.9112 & 0.1152 & 8.35  & 0.29 & 0.1013  \\
&0.925 & 0.9398 & 0.0857 & 13.69 & 0.19 & 0.1581  \\
&0.950 & 0.9508 & 0.0857 & 15.85 & 0.19 & 0.1810  \\ \hline
\multirow{7}{*}{\begin{tabular}[c]{c}SG\end{tabular}}
&0.800 & 0.8957 & 0.3337 & 14.62 & 0.30 & 0.2548  \\
&0.825 & 0.9201 & 0.3284 & 16.90 & 0.27 & 0.2811  \\
&0.850 & 0.9417 & 0.3271 & 19.81 & 0.34 & 0.3168  \\
&0.875 & 0.9581 & 0.3259 & 25.54 & 0.42 & 0.3973  \\
&0.900 & 0.9660 & 0.3107 & 30.54 & 0.54 & 0.4578  \\
&0.925 & 0.9694 & 0.2530 & 33.21 & 0.42 & 0.4577  \\
&0.950 & 0.9708 & 0.2469 & 35.38 & 0.38 & 0.4835  \\
\hline
\end{tabular}}
\end{table}

 In this subsection, we compare {\em r}-HUMO with the active learning based (ACTL) alternative. We have implemented of both of the techniques proposed in \cite{arasu2010active} and \cite{bellare2012active} respectively. Our experiments showed that they perform similarly on the achieved quality and required manual work. Here, we present the comparative evaluation results between {\em r}-HUMO and the technique proposed in \cite{arasu2010active}. As \cite{arasu2010active}, we employ Jaccard similarity, edit distance and number similarity on attributes used in Subsection~\ref{sec:setup} as the similarity space for ACTL. On DS, the used attributes are {\em title} and {\em authors}; on AB, they are {\em product name} and {\em product description}; and on SG, they are \emph{song title}, \emph{release information}, \emph{artist name}, \emph{duration}, \emph{artist familiarity}, \emph{artist hotness} and \emph{year}. ACTL uses sampling to estimate the achieved precision level of a given classification solution; therefore it also requires manual work.

  In our experiments, the required precision and recall levels are set to be the same for {\em r}-HUMO. Considering that ACTL can not enforce recall level; at each given precision level, we record {\em r}-HUMO and ACTL's performance difference on the achieved recall and the consumed human cost. The detailed comparison results between {\em r}-HUMO and ACTL are presented in Table~\ref{tab:compare-act}, in which $\psi$ represents the percentage of manual work, and $\Delta$ denotes the performance difference between the two methods on a specified metric. It can be observed that the achieved recall level by ACTL generally decreases with the specified precision level. In all the test cases, {\em r}-HUMO achieves higher recall levels than ACTL. We also record the additional human cost required by {\em r}-HUMO for the absolute recall improvement of $1\%$ over ACTL (at the last columns of Table~\ref{tab:compare-act}. It can be observed that, with both precision and recall set at the high level of 0.95, the cost is as low as 0.1762\% on DS, 0.1810\% on AB and 0.4835\% on SG.

\begin{table}[!h]
\centering
\caption{{\em r}-HUMO vs ACTL on F1 given the Same Precision.}
\vspace{-0.05in}
\label{tab:compare-act-f1}
\setlength{\tabcolsep}{0.8mm}{
\begin{tabular}{|c|c|c|c|c|c|c|}
\hline
\multirow{2}{*}{\begin{tabular}[c]{c}Data\\set\end{tabular}} &\multirow{2}{*}{\begin{tabular}[c]{c}Required\\Precision\end{tabular}} & \multicolumn{2}{c|}{Achieved F1} & \multicolumn{2}{c|}{$\psi (\%)$} & \multirow{2}{*}{$\frac{\Delta \psi}{100 \cdot \Delta F1}$}\\
\cline{3-6}
& & {\em r}-HUMO & ACTL & {\em r}-HUMO & ACTL & \\
\hline
\multirow{6}{*}{\begin{tabular}[c]{c}DS\end{tabular}}
&0.825 & 0.8758 & 0.8057 & 4.97  & 3.46 & 0.2154  \\
&0.850 & 0.8872 & 0.8067 & 5.08  & 3.70 & 0.1714  \\
&0.875 & 0.9013 & 0.8130 & 5.23  & 2.98 & 0.2548  \\
&0.900 & 0.9199 & 0.8187 & 5.47  & 3.17 & 0.2273  \\
&0.925 & 0.9444 & 0.8220 & 5.90  & 3.83 & 0.1691  \\
&0.950 & 0.9688 & 0.8161 & 6.71  & 2.56 & 0.2718  \\ \hline
\multirow{7}{*}{\begin{tabular}[c]{c}AB\end{tabular}}
&0.800 & 0.9055 & 0.2626 & 6.26  & 0.29 & 0.0929  \\
&0.825 & 0.9080 & 0.2408 & 6.75  & 0.28 & 0.0970  \\
&0.850 & 0.9154 & 0.2653 & 7.23  & 0.28 & 0.1069  \\
&0.875 & 0.9267 & 0.2053 & 7.60  & 0.27 & 0.1016  \\
&0.900 & 0.9374 & 0.2053 & 8.35  & 0.29 & 0.1101  \\
&0.925 & 0.9533 & 0.1575 & 13.69 & 0.19 & 0.1696  \\
&0.950 & 0.9600 & 0.1575 & 15.85 & 0.19 & 0.1951  \\ \hline
\multirow{7}{*}{\begin{tabular}[c]{c}SG\end{tabular}}
&0.800 & 0.9297 & 0.4807 & 14.62 & 0.30 & 0.3189  \\
&0.825 & 0.9430 & 0.4790 & 16.90 & 0.27 & 0.3584  \\
&0.850 & 0.9546 & 0.4780 & 19.81 & 0.34 & 0.4085  \\
&0.875 & 0.9632 & 0.4769 & 25.54 & 0.42 & 0.5166  \\
&0.900 & 0.9673 & 0.4629 & 30.54 & 0.54 & 0.5948  \\
&0.925 & 0.9690 & 0.4013 & 33.21 & 0.42 & 0.5776  \\
&0.950 & 0.9727 & 0.3939 & 35.38 & 0.38 & 0.6047  \\
\hline
\end{tabular}}
\end{table}

 It can be observed that given the same precision requirement, ACTL and {\em r}-HUMO might actually achieve different precision levels. Therefore, we also compare their actual performance on the F1 metric and record the additional human cost required by {\em r}-HUMO for the absolute F1 improvement of $1\%$ over ACTL. The detailed results are presented in Table~\ref{tab:compare-act-f1}. Similar to what was observed in Table~\ref{tab:compare-act}, the additional human cost generally increases with the specified precision level. On DS, the additional human cost of {\em r}-HUMO for 1\% increase in F1 score is maxed at 0.2718\%. On AB and SG, it is as low as 0.1951\% and 0.6047\% respectively. Along with the results presented in Table~\ref{tab:compare-act}, these results clearly demonstrate that compared with ACTL, {\em r}-HUMO can effectively improve the resolution quality with reasonable ROI in terms of human cost.

\subsection{{\em r}-HUMO: Real-time vs Batch Mode} \label{sec:batch}

    In this subsection, we compare the performance of {\em r}-HUMO in the real-time and batch-mode settings. Given the same GPR approximations, we compare the performance by the frequency of required human and machine interaction and the total amount of required manual work (the size of $D_H$ excluding sampled pairs) as well as the achieved quality. The detailed evaluation results on DS and AB are presented in Table~\ref{tb:bm} and Table~\ref{tb:bmquality}. The results on SG are similar, therefore omitted here due to space limit. It can be observed that the batch mode achieves very similar performance to the real-time mode in terms of resolution quality and human cost, while significantly reducing the frequency of required interactions (up to 90+\%). These experimental results clearly validate the efficacy of the proposed batch mechanism.

\begin{table}[!ht]
\centering
\caption{{\em r}-HUMO: Real-time vs Batch on Achieved Quality.}
\vspace{-0.05in}
\label{tb:bmquality}
\setlength{\tabcolsep}{2.4mm}{
\begin{tabular}{|c|c|cc|cc|}
\hline
\multirow{3}{*}{\begin{tabular}[c]{c}Dataset\end{tabular}} &\multirow{2}{*}{\begin{tabular}[c]{c}Required\\Quality\end{tabular}}& \multicolumn{4}{c|}{Achieved Quality} \\ \cline{3-6}
& &\multicolumn{2}{c|}{Real Time} &\multicolumn{2}{c|}{Batch Manner} \\ \cline{2-6}
&$\alpha$=$\beta$     & $\alpha$ & $\beta$      & $\alpha$ & $\beta$      \\ \hline
\multirow{6}{*}{\begin{tabular}[c]{c}DS\end{tabular}}
&0.825    & 0.8986 & 0.8537 & 0.8986 & 0.8537     \\
&0.850    & 0.9009 & 0.8759 & 0.9009 & 0.8759     \\
&0.875    & 0.9058 & 0.9008 & 0.9059 & 0.9010     \\
&0.900    & 0.9253 & 0.9257 & 0.9256 & 0.9261     \\
&0.925    & 0.9503 & 0.9506 & 0.9511 & 0.9505     \\
&0.950    & 0.9755 & 0.9694 & 0.9766 & 0.9692     \\ \hline
\multirow{7}{*}{\begin{tabular}[c]{c}AB\end{tabular}}
&0.800    & 0.9620 & 0.8540 & 0.9619 & 0.8533     \\
&0.825    & 0.9622 & 0.8587 & 0.9622 & 0.8584     \\
&0.850    & 0.9627 & 0.8712 & 0.9627 & 0.8713     \\
&0.875    & 0.9635 & 0.8908 & 0.9635 & 0.8909     \\
&0.900    & 0.9643 & 0.9105 & 0.9643 & 0.9106     \\
&0.925    & 0.9671 & 0.9398 & 0.9671 & 0.9398     \\
&0.950    & 0.9693 & 0.9508 & 0.9693 & 0.9508     \\ \hline
\end{tabular}}
\end{table}

\vspace{-0.13in}

\begin{table}[!ht]
\centering
\caption{{\em r}-HUMO: Real-time vs Batch on Human Cost.}
\vspace{-0.05in}
\label{tb:bm}
\setlength{\tabcolsep}{0.75mm}{
\begin{tabular}{|c|c|cc|c|}
\hline
\multirow{2}{*}{\begin{tabular}[c]{c}Data\\set\end{tabular}} &Required Quality & \multicolumn{2}{c|}{Size of $D_H$ (excl. samples)}  & Interaction Frequency   \\ \cline{2-4}
&$\alpha$=$\beta$ & Real Time & Batch & Reduction(\%) \\ \hline
\multirow{6}{*}{\begin{tabular}[c]{c}DS\end{tabular}}
&0.825   & \textbf{24}    & \textbf{24}                & 90.17   \\
&0.850   & \textbf{150}   & \textbf{150}               & 91.10   \\
&0.875   & \textbf{325}   & 326                        & 91.28   \\
&0.900   & \textbf{638}   & 640                        & 91.26   \\
&0.925   & \textbf{1132}  & 1134                       & 91.04   \\
&0.950   & \textbf{2074}  & 2118                       & 91.15   \\ \hline
\multirow{7}{*}{\begin{tabular}[c]{c}AB\end{tabular}}
&0.800   & \textbf{4049}  & 4089                       & 83.19   \\
&0.825   & \textbf{5646}  & 5668                       & 81.27   \\
&0.850   & 7195           & \textbf{7156}              & 78.09   \\
&0.875   & \textbf{8323}  & 8335                       & 74.50   \\
&0.900   & \textbf{10410} & 10427                      & 70.04   \\
&0.925   & \textbf{25615} & 25634                      & 57.35   \\
&0.950   & \textbf{34547} & 34557                      & 44.59   \\ \hline
\end{tabular}}
\end{table}

\vspace{-0.15in}

\subsection{Efficiency and Scalability} \label{sec:scalability}

  In this subsection, we evaluate the efficiency and scalability of our {\em r}-HUMO implementation on different data scales. We perform random sampling on the DS dataset to generate the test workloads with different data scales. We measure the efficiency by the consumed run time.

\begin{figure}[!h]
\centering
\includegraphics[width=0.7\linewidth]{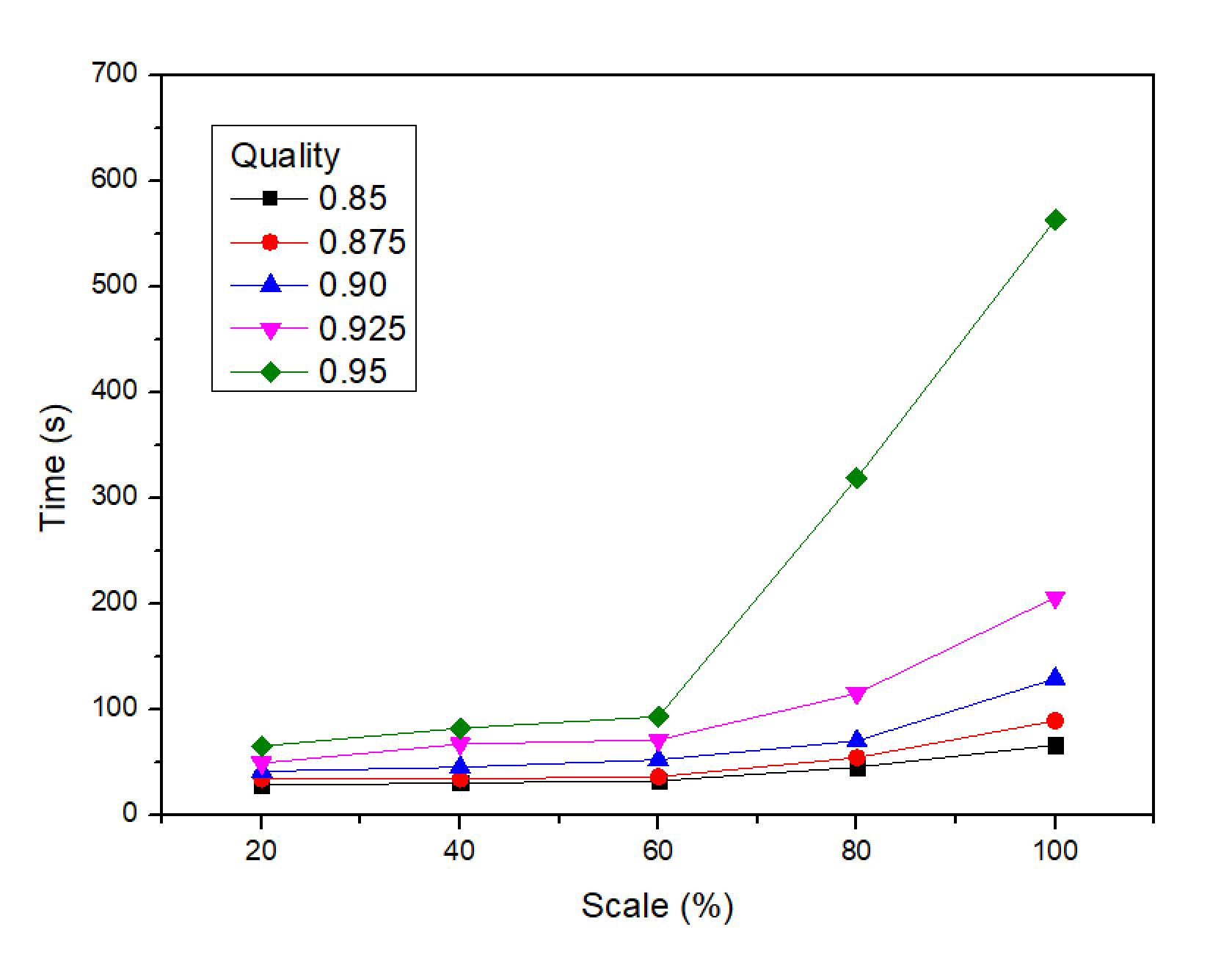}
\caption{Evaluation of {\em r}-HUMO Efficiency and Scalability on DS.}
\label{fig:efficiency}
\end{figure}

  The evaluation results are presented in Figure.~\ref{fig:efficiency}. It can be observed that as data scale increases, the run time increases polynomially as dictated by the complexity analysis results. As expected, the run time increases more dramatically with data scale as the quality requirement becomes stricter.
	
